\documentclass[a4,aps,pre,twocolumn,amsmath,superscriptaddress,amssymb,showpacs]{revtex4}
\usepackage{graphicx} 
\usepackage{dcolumn} 
\usepackage{bm} 
\usepackage[english]{babel}

\begin{document}

\title{Dimer diffusion in a washboard potential }

\author{E. Heinsalu}
  \affiliation{Institute of Theoretical Physics, University of Tartu,
  T\"ahe 4, 51010 Tartu, Estonia}

\author{M. Patriarca}
  \affiliation{Institute of Theoretical Physics, University of Tartu,
  T\"ahe 4, 51010 Tartu, Estonia}

\author{F. Marchesoni}
   \affiliation{Dipartimento di Fisica, Universit\`a di Camerino, I-62032 Camerino, Italy}


%
%

\begin{abstract}
The transport of a dimer, consisting of two Brownian particles
bounded by a harmonic potential, moving on a periodic substrate is
investigated both numerically and analytically. The mobility and
diffusion of the dimer center of mass present distinct properties
when compared with those of a monomer under the same transport conditions.
Both the average current and the diffusion coefficient are found
to be complicated non-monotonic functions of the driving force.
The influence of dimer equilibrium length, coupling strength and
damping constant on the dimer transport properties are also
examined in detail.\\

\end{abstract}
\pacs{05.60.-k, 05.40.-a, 68.43.Mn} \maketitle

\section{Introduction}
\label{introduction}

One particular example of Brownian motion on a periodic substrate is
the diffusion of atoms and molecules on crystal surfaces
\cite{frenken1985}. This mechanism is of both conceptual and
technological interest \cite{naumovets1985}, being relevant to
heterogeneous nucleation, catalysis, surface coating, thin-film
growth, etc. Individual atoms diffusing on a surface can
eventually meet and form dimers or trimers. For example, on the
semiconductor Si(100) or Ge(100) surface, most of the deposited Si
or Ge atoms form dimers. Atoms adsorbed on metal surfaces may also
form closely packed islands that diffuse as a whole
\cite{PortoLett2000, wang1990, voter1992}. This raises the issue
of the role of the internal degrees of freedom on the transport of
extended objects through micro- and submicro-devices.

One of the most important problems in modern nanotechnology is how
to manipulate small particles in order to perform a preassigned
operation. For instance, the mobility and diffusivity of atoms
adsorbed onto crystal surfaces can be controlled by applying
deterministic forces \cite{naumovets2002, alanissila2002}. A direct
manipulation method consists in applying a constant direct
current (dc) local electric field by means of a scanning tunnel
microscope tip \cite{swartzentruber1996}. A selected adatom or
admolecule with nonzero charge will then move in the direction of
the electric force; neutral particles will be forced into a region
of a stronger field due to induced polarization \cite{riken}. This
problem can be modeled as a Brownian motion on a tilted periodic
two-dimensional (2D) substrate.

In this work we study the transport of a dimer confined on a
periodic substrate with a focus on the effects of the internal degrees
of freedom on its mobility and diffusivity. For simplicity, we
restrict our analysis to substrates in two or higher dimensions, which
can be effectively reduced to one-dimensional (1D) systems. In the
simple case of a dimer driven by a constant force oriented along a
symmetry axis of a 2D substrate, one wants to characterize the
stationary transport in the force direction, whereas transverse
diffusion is not affected by the bias;
for a full 2D treatment, see, e.g., Ref.~\cite{romero}.
Of course, the results of the
present paper apply well also to a variety of physical and
biological systems, where the particle dynamics is naturally
constrained to (quasi-)1D substrates. Examples of current interest
include colloids \cite{colloids} or cold atoms \cite{renzoni} in
optical traps, superconducting vortices in lithographed tracks
\cite{vortices}, ion-channels \cite{ions}, cell membranes
\cite{cells}, artificial and natural nanopores \cite{pores}, etc.

This paper is organized as follows: In Sec.~\ref{model} we introduce
the model, define the units and give the details of our numerical
simulations. Our numerical results are presented in
Sec.~\ref{results}. In particular, the role of the dimer length in
the transport properties is studied in Sec.~\ref{length}; the
monomer like regimes (for weak and strong couplings)
are discussed in Sec.~\ref{tilting}; finally, the influence of the
coupling strength and of the damping constant on the dimer transport
are analyzed in Sec.~\ref{coupling}. Potential applications of our
results to 1D irreducible devices are sketched in
Sec.~\ref{conclusion}.

\section{Model}
\label{model}

A monomer moving on a 1D periodic substrate with potential
$U_0(x)=U_0(x+L)$ under the influence of an external dc bias $F$
and at finite temperature $T$ can be described by the Langevin
equation (LE),
\begin{eqnarray} \label{monomer}
m \ddot{x} =  - \eta \dot{x} - \frac{d U_0(x)}{d x} + F + \xi (t) \,
.
\end{eqnarray}
Here $\eta = m \gamma $ is the viscous friction coefficient, with
$\gamma $ being a damping constant and $m$ the mass of the
Brownian particle. The stochastic force $\xi (t)$ represents the
environmental fluctuations and is modeled by a Gaussian white
noise with zero mean, $\langle \xi (t) \rangle = 0$, and
auto-correlation function (ACF),
\begin{equation}
\langle \xi (t) \, \xi (t') \rangle = 2 \, \eta k_{B} T \,
\delta(t-t') \, .
\end{equation}

For a symmetric dimer the corresponding LE's have the form
\begin{eqnarray} \label{dimer}
m\ddot{x}_1 &=& - \eta
\dot{x}_1 - \frac{\partial U(x_1, x_2)}{\partial x_1} + F + \xi_1 (t) \, , \nonumber \\
m\ddot{x}_2 &=& - \eta \dot{x}_2 - \frac{\partial U(x_1,
x_2)}{\partial x_2}+ F  + \xi_2 (t) \, ,
\end{eqnarray}
where $\xi _{i}(t)$, $i=1,2$, are two independent zero-mean
stochastic processes with ACF,
\begin{equation}
\langle \xi_i (t) \xi_j (t') \rangle = 2 \eta k_B T \, \delta_{ij}
\, \delta (t-t') \, .
\end{equation}
Note that the inter-particle interaction is incorporated in the
potential function,
\begin{equation} \label{total}
U(x_1, x_2) = U_0(x_1) + U_0(x_2) + \frac{K}{2} (x_2 - x_1 -
a_0)^2 \, .
\end{equation}
That is, we assume the interaction between the two dimer particles
to be harmonic with coupling constant $K$ and equilibrium distance
$a_0$. The simplest choice for the periodic substrate potential is
\cite{risken},
\begin{equation} \label{Mperpot}
U_0(x) = A_0 \cos (kx) \, ,
\end{equation}
with $k=2\pi /L$.

The LE's (\ref{monomer}) and (\ref{dimer}) can be conveniently
rescaled. By introducing suitable space, energy, and time units,
\begin{equation}
\lambda = 1/k \, , \quad \epsilon = A_0 \, , \quad \tau =
\sqrt{\lambda ^2 m /\epsilon} \, ,
\end{equation}
we define the dimensionless quantities:
\begin{eqnarray} \label{resc}
\tilde{x}=\frac{x}{\lambda } \, , \quad
\tilde{a}_0=\frac{a_0}{\lambda } \, , \quad \tilde{T}=\frac{k_B
T}{\epsilon } \, , \quad \tilde{F} =\frac{ \lambda }{
\epsilon }F \, , \nonumber \\
\tilde{K} = \frac{ \lambda ^2}{\epsilon } K \, , \quad \tilde{t} =
\frac{t}{\tau } \, , \quad \tilde{\gamma } = \gamma \tau \, , \quad
\tilde{ \xi }(\tilde{t}) = \frac{ \lambda }{\epsilon } \xi(t) \, .
\end{eqnarray}
No particle can be trapped by the potential (\ref{Mperpot}) under
any circumstances for tilting larger than the critical value
$\tilde{F}_{\mathrm{cr}}=1$ (in rescaled units). In the following we
drop the tilde altogether.

After rescaling, the LE (\ref{monomer}) for a monomer moving in
the potential (\ref{Mperpot}) reads,
\begin{equation} \label{monomerDL}
\ddot{x} = - \gamma \dot{x} + \sin{x} + F + \xi (t) \, ,
\end{equation}
where the ACF of the rescaled noise is $ \langle \xi (t) \, \xi (t')
\rangle = 2 \, \gamma T \, \delta(t-t') $. Analogously, the coupled
LE's (\ref{dimer}) for a symmetric harmonic dimer in the same
substrate potential become [see Eq.~(\ref{total})],
\begin{eqnarray} \label{dimerDL}
\ddot{x}_1 &=& - \gamma \dot{x}_1 + \sin{x_1} + F + K(x_2 - x_1 -a_0) + \xi_1 (t) \, , \nonumber \\
\ddot{x}_2 &=& - \gamma \dot{x}_2 + \sin{x_2} + F - K(x_2 - x_1
-a_0) + \xi_2 (t) \, ,  \nonumber \\
\end{eqnarray}
with $ \langle \xi _i (t) \, \xi _j (t') \rangle = 2 \, \gamma T \,
\delta _{ij} \delta(t-t')$.

The dimensionless LE (\ref{monomerDL}) for a monomer and
(\ref{dimerDL}) for a dimer have been integrated numerically
through a standard Milstein algorithm~\cite{milstein}. Individual stochastic
trajectories were simulated for different time lengths $t_{\rm
max}$ and time steps $\Delta t$, so as to ensure appropriate
numerical accuracy. Average quantities have been obtained as
ensemble averages over $10^4$ trajectories; transient effects
have been estimated and subtracted.

\section{Results: Mobility and diffusion}
\label{results}

When considering a pair of interacting Brownian particles, it is
natural to study the motion of their center of mass,
\begin{equation}
X = \frac{1}{2}(x_1+x_2) \, .
\end{equation}
The quantities that best characterize the stationary dimer flow
are: (a) the net velocity,
\begin{equation}
v = \lim _{t \rightarrow \infty } \frac{\langle X(t) \rangle }{t},
\end{equation}
or, equivalently, the related mobility, $\mu = v/F$; (b) the
diffusion coefficient,
\begin{equation}
D = \lim _{t \rightarrow \infty } \frac{\langle \delta X^2(t)
\rangle }{2t} \, ,
\end{equation}
where $\langle \delta X^2 \rangle$ is the mean square displacement
of the center of mass, i.e.,
\begin{eqnarray}
\langle \delta X^2 \rangle &=& \langle X^2 \rangle - \langle X
\rangle ^2  \\
&=& \frac{1}{4} \langle \delta x_1^2 \rangle +  \frac{1}{4}
\langle \delta x_2^2 \rangle + \frac{1}{2}(\langle x_1 x_2 \rangle
- \langle x_1 \rangle \langle x_2 \rangle). \nonumber
\label{deltaX}
\end{eqnarray}

For the following discussion we also introduce the relative
coordinate $Y$,
\begin{equation}
Y=x_2-x_1 \, ,
\end{equation}
representing the dimer size. 
The quantity $Y$ can in principle also become also
negative. However, this happens only when the dimer oscillations around the
equilibrium position become very large. In the range of parameters adopted in the
present paper, we have verified that $Y$ remains positive even for small values
of the elastic constant $K$, where one recovers the monomer limit. In fact, the distance
$Y$ can become negative if both monomers fall into the same valley. In our
simulations, the dimer length (at rest) varies in the range $a_0 \in [ L, 2 L ]$. Thus, the
monomers start out in different potential valleys and are observed to stay so for all
times (i.e., configurations with $Y < 0$ do not occur).

The LE's (\ref{dimer}) can be
rewritten as a LE for the center of mass coordinate $X$ and one for
the dimer length $Y$, that is,
\begin{eqnarray}
\ddot{X} &=& - \gamma \dot{X} + \cos{(Y/2)}\sin{X} + F + Q (t)/\sqrt{2} \, , \label{CMR2a} \\
\ddot{Y} &=& - \gamma \dot{Y} + 2\cos{X} \sin{(Y/2)} - 2K(Y -a_0)
+ \sqrt{2} q (t) \, . \nonumber \\ \label{CMR2b}
\end{eqnarray}
Note that the two noises $Q(t)=[\xi _1(t)+\xi _2(t)]/\sqrt{2}$ and
$q(t)=[\xi _2(t)-\xi _1(t)]/\sqrt{2}$ are uncorrelated and have
the same statistics as $\xi_{1,2}(t)$, namely, $\langle q(t)
\rangle = \langle Q(t) \rangle =0$ and
\begin{equation}
\langle q(t) q(t') \rangle = \langle Q(t) Q(t') \rangle = 2 \gamma T
\delta (t-t') \, .
\end{equation}
In the absence of a substrate potential the mobility of both a
monomer and a dimer is $\mu _0=1/\gamma $. Correspondingly, the
free diffusion coefficient for a monomer, $D_0(T) \equiv T/\gamma
$, is twice as large as that for a dimer, $D_0(T/2)$.

\begin{figure}[t]
\begin{center}
\includegraphics[width=3in]{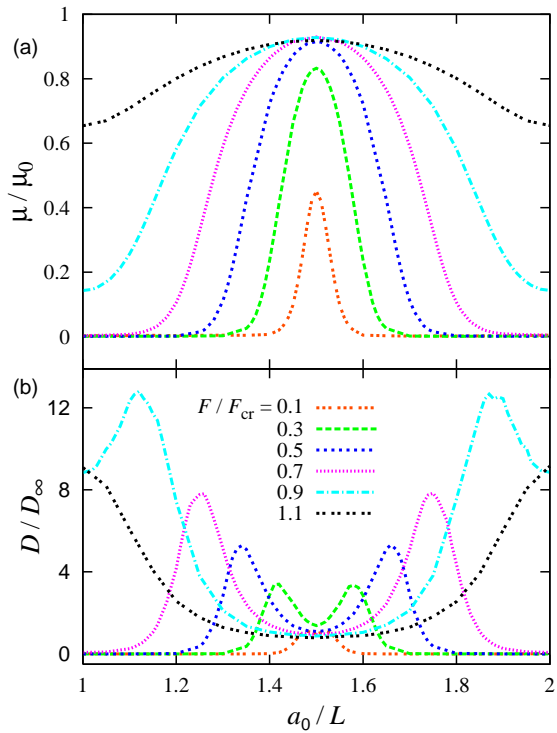}
\caption{(Color online) Mobility (a) and diffusion coefficient (b)
versus the dimer length $a_0$ for different values of the tilting
force $F$. Simulation parameters: coupling constant $K = 1.5$,
temperature $T=0.1$ and $\gamma =1$. $D_{\infty}$ is the free
dimer diffusion coefficient $D_{\infty} = D_0(T/2)$; see text. }
\label{dep-aF}
\end{center}
\end{figure}
%

\subsection{The role of the dimer length }
\label{length}

At variance with a monomer, a dimer has two degrees of freedom. This
affects its diffusion dynamics \cite{braun2001PRE}  to the point
that its diffusion coefficient $D$ can develop a non-monotonic
dependence on the dimer parameters. For instance, dimer transport
strongly depends on the ratio between the period $L$ of the
substrate and the natural length $a_0$ of the dimer
\cite{fusco2003TSF,brown1990SS, patriarca2005}.

In the absence of an external force, $F=0$, at low temperature the
diffusion coefficient of a rigid dimer decreases monotonically on
raising the dimer length $a_0$ from $L/2$ to $L$. This can be well
understood from Eq.~(\ref{CMR2a}). In the limit $K=\infty $ the
dimer length is just $Y=a_0$ and the force $\cos{(Y/2)\sin{X}}$
acting on $X(t)$, Eq.~(\ref{CMR2a}), corresponds to a periodic
potential with amplitude $|\cos{(Y/2)}|$. For $a_0=L/2=\pi $ this
quantity is zero and the dimer center of mass undergoes free
diffusion. For $a_0=L=2\pi $ the periodic potential amplitude is
maximum, $|\cos(Y/2)|=1$; diffusion in a periodic potential is known
to be suppressed compared to free diffusion \cite{lifson1962,
festa}. Therefore, the maxima and minima of $D$ versus $a_0$
coincide with the minima and the maxima of the modulating factor
$|\cos(a_0/2)|$, respectively. This conclusion applies also to the
case of finite elastic constants as long as $\langle Y(t) \rangle
\approx a_0$, that is for rigid dimers, $K \gg 1 $, at low
temperatures, $T\ll 1$. (For the opposite limit of weak dimers,
$K\ll 1$, see Sec.~\ref{tilting}.)

\begin{figure}[t]
\begin{center}
\includegraphics[width=3in]{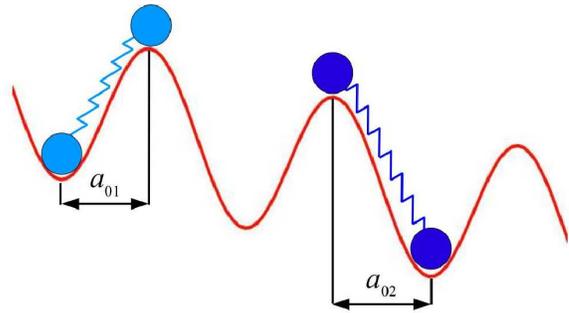}
\caption{(Color online) Dimer configurations corresponding to
zero pinning force and maximum diffusion coefficient; 
see also Fig.~\ref{dep-aF} and text.}
\label{minmax}
\end{center}
\end{figure}

In the presence of a sub-threshold external force, $F<F_\mathrm{cr}$,
the diffusion coefficient $D$ is a nonmonotonic function of the
dimer length $a_0$, as shown in Fig.~\ref{dep-aF}(b).  The numerical
results in Fig.~\ref{dep-aF} have been obtained by simulating a
relatively rigid, $K=1.5$, and moderately damped, $\gamma=1$, dimer.
In the case of a strong to moderately damped {\it monomer} in a
washboard potential, the curves $D(F,T)$ are known to develop a peak
around $F_\mathrm{cr}$, where the barrier height of the tilted periodic
potential $U_0(x)-Fx$ vanishes \cite{costantini}. Analogously, in
the case of a dimer, $D$ attains a maximum for dimer lengths such
that the effective pinning force also vanishes, i.e., for $a_0$
equal to the distances between maxima and minima of the washboard
potential, see Fig.~\ref{minmax}. 
In the case of a driven rigid dimer with $F < F_\mathrm{cr}=1$, 
this takes place for equilibrium lengths
$a^{\pm}_{0} = (L/2)[1 \pm (2/\pi) \arcsin(F)]$. Note that
$a_0^{\pm}$ are given $\mathrm{mod}(L)$ and $a_0^+ + a_0^-=L$.

Figure~\ref{dep-aF}(a) demonstrates that the mobility is
smallest for commensurate dimers with $a_0=L$ and largest for
$a_0=L/2$ (see also Ref.~\cite{patriarca2005}). The smaller the
applied constant force, the smaller is the $a_0$ range around
$a_0=L/2$, where the mobility of the dimer is significantly
different from zero. For large enough tilting the dimer is
considerably mobile, no matter what is the value of $a_0$. For $F\to\infty$ the
mobility $\mu \to \mu _0 $ and the effective diffusion coefficient
$D\to D_{\infty}=D_0(T/2)$. We remark that the $a_0$ dependencies of
$\mu$ and $D$ shown in Fig.~\ref{dep-aF} are given
$\mathrm{mod}(L)$ \cite{patriarca2005}.
In fact, the system dynamics, as given by Eqs.~(\ref{dimerDL}), 
is invariant under the change $a_0 \to a_0 + L$ and $x_1 \to x_1 - L$
(or $x_2 \to x_2 + L$).

\begin{figure}[t]
\begin{center}
\includegraphics[width=3.5in]{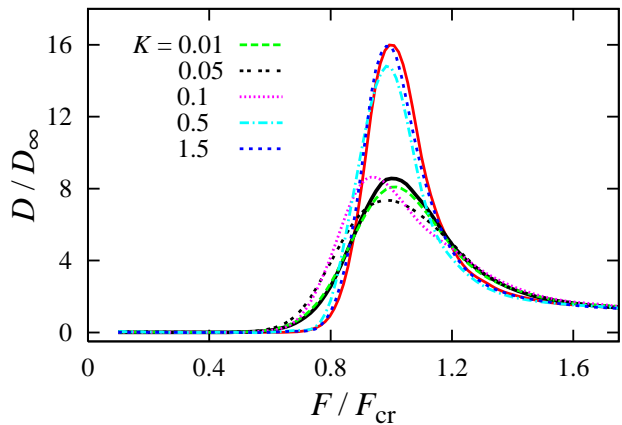}
\caption{(Color online) Diffusion coefficient $D$ versus the
tilting force $F$ for a dimer length $a_0/L = 1$ and different
coupling constants $K$; $T=0.1$ and $\gamma =1$. The corresponding
curves for monomers of temperature $T$ (solid, black) and $T/2$
(solid, red) are drawn for comparison (see text).}
\label{dep-FK-a1}
\end{center}
\end{figure}
%

\subsection{Monomer like regimes }
\label{tilting}

In the case of a monomer the mobility and the diffusion in a
tilted periodic potential are in general well understood. In the
low-temperature regime, $T \ll 1$, the particle mobility is close
to zero for sub-threshold tilting (locked state). Around a
depinning threshold $F_d$, the mobility  grows sharply and in the
large force limit reaches the free particle limit $\mu _0$
(running state). If the temperature is increased, the transition
from the locked to the running state is smoother. In the
overdamped regime, convenient fully analytical expressions are
available for both the mobility $\mu(F,T)$ (the Stratonovich formula
\cite{stratonovich}) and the diffusion coefficient $D(F,T)$ (the Cox
formula \cite{reimann2001}). For small biases and low temperature,
the diffusion coefficient is suppressed compared to the free
diffusion $D_0(T)$; in linear response theory $D(F,T) \simeq
\mu(F,T)T$ \cite{risken}. Depinning occurs around the critical
tilt, i.e., $F_d\simeq F_\mathrm{cr}$, as signaled by $D(F,T)$
overshooting $D_0$ \cite{costantini}; the lower the temperature,
the more prominent is the growth of the depinning diffusion peak.
In the large force limit, the free diffusion regime $D_0(T)$ is
eventually recovered.

In the underdamped limit, $\gamma \ll \sqrt{F_\mathrm{cr}}$, the
mobility and the diffusion coefficient display a similar behavior
with one significant difference: the depinning threshold $F_d$ is
a monotonic function of the damping constant with
\begin{equation}
\label{lowg} \lim_{\gamma \to 0}F_d \simeq
3.36\gamma\sqrt{F_\mathrm{cr}}
\end{equation}
and $F_d\simeq F_\mathrm{cr}$ for $\gamma \gtrsim
\sqrt{F_\mathrm{cr}}$ \cite{risken,washboard}.

\begin{figure}[t]
\begin{center}
\includegraphics[width=3in]{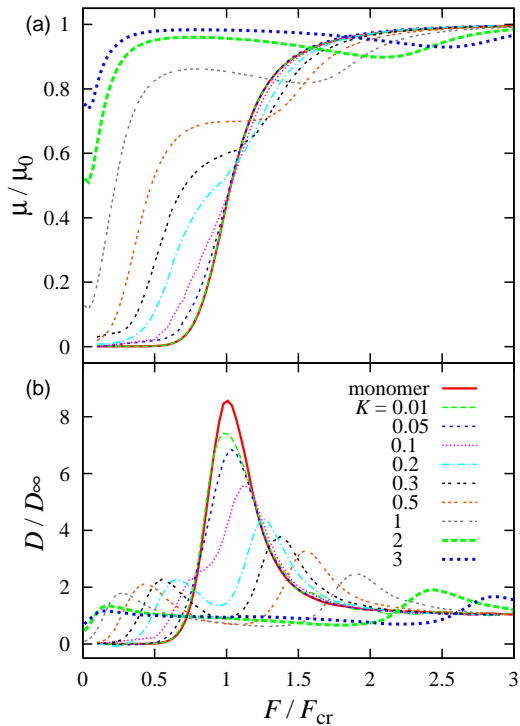}
\caption{(Color online) Mobility (a) and diffusion coefficient (b)
versus the tilting force $F$ for an equilibrium distance $a_0/L =
1.5$ and for different values of the coupling constant $K$;
$T=0.1$ and $\gamma =1$. In both panels the results are compared
with the corresponding monomer curves (see text).} \label{dep-FK}
\end{center}
\end{figure}

In the case of a dimer, the general behavior recalls that of a
monomer, namely, both the transition of the rescaled mobility from
$0$ to $\mu _0$ and the corresponding enhancement of the diffusion
coefficient above its free diffusion value still occur as the
tilting force is increased past the depinning threshold. The monomer
dynamics is a useful benchmark to check the accuracy of our
simulations for the dimer diffusion. Indeed, in the limit $K \to 0
$, Eq.~(\ref{deltaX}) boils down to $\langle \delta X^2 \rangle =
\langle \delta x_1^2 \rangle/2$, with $x_1$ obeying the monomer LE
(\ref{monomerDL}) with temperature $T$. It follows that for a weak
dimer, $K \ll 1$, the ratio $D/D_{\infty}$ is closely reproduced by
the analytical curve $D(F,T)/D_0(T)$ obtained from the monomer LE
(\ref{monomerDL}). This argument applies to both commensurate,
Fig.~\ref{dep-FK-a1}, and incommensurate dimers,
Fig.~\ref{dep-FK}(b).

Rigid dimers also behave like monomers. In the limit $K \to \infty
$, the solution of Eq.~(\ref{CMR2b}) is $Y(t) \equiv Y=  a_0 $ and
Eq.~(\ref{CMR2a}) is then equivalent to the monomer LE
(\ref{monomerDL}) with temperature $T/2$ and substrate amplitude
(critical tilt) $\cos(a_0/2)$. Accordingly, for commensurate dimers
with $a_0$ equal to an integer multiple of the substrate constant
$L$, the ratio $D/D_{\infty}$ is reproduced by the curve
$D(F,T/2)/D_0(T/2)$ obtained for a monomer on a tilted cosine
potential with amplitude $|\cos (a_0/2)| =1$ and temperature $T/2$
(see Fig.~\ref{dep-FK-a1}).

Note that for large values of damping the monomer curve can also be computed
analytically through the Cox formula \cite{reimann2001}. The data in
Fig.~\ref{dep-FK-a1} confirm that for increasingly large $K$ the
depinning threshold approaches $F_{\mathrm{cr}}=1$ from below, as
the effective critical tilt $\langle |\cos(\psi/2)| \rangle$ tends to
unity. Not surprisingly, for the commensurate dimer of
Fig.~\ref{dep-FK-a1} the mobility curve coincides with the monomer
mobility $\mu(F,T)$, in the weak coupling limit, and with the
monomer mobility at half the temperature $T$, $\mu(F,T/2)$, in the
strong coupling limit; both limiting curves are closely approximated
by the Stratonovich formula (not shown).

For $K\to \infty $ incommensurate dimers behave like monomers
moving on a tilted cosine potential with amplitude $|\cos (a_0/2)|
<1$ and temperature $T/2$ (see also Fig.~\ref{dep-Fa} for a finite
coupling). When $a_0$ is equal to a half-integer multiple of the
substrate constant $L$, the amplitude of the effective substrate
acting on the dimer coordinate $X$ vanishes, $|\cos (a_0/2)|=0$, 
and the dimer diffusion becomes insensitive to the substrate, with
mobility $\mu_0$ and diffusion coefficient $D_0(T/2)$.

Figures~\ref{dep-FK} and \ref{dep-Fa} indicate that for a finite $K$
the dimers exhibit a much more complicated behavior, which will be
discussed in the forthcoming section.

\subsection{The dependence on the coupling strength }
\label{coupling}

\begin{figure}[t]
\begin{center}
\includegraphics[width=3in]{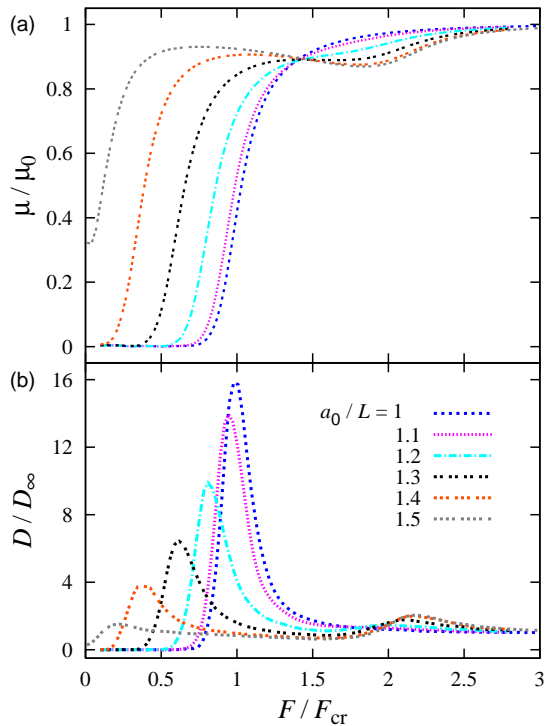}
\caption{(Color online) Mobility (a) and diffusion coefficient (b)
versus the tilting force $F$ for a coupling constant $K = 1.5$ and
for different values of the equilibrium distance $a_0$. $T$ and
$\gamma $ have the same values as in Figs.~\ref{dep-FK} and
\ref{dep-FK-a1}. } \label{dep-Fa}
\end{center}
\end{figure}

\begin{figure}[t]
\begin{center}
\includegraphics[width=3in]{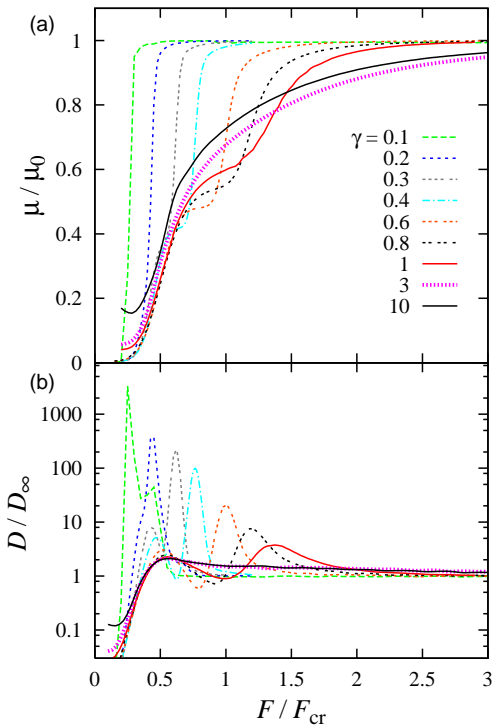}
\caption{(Color online) Mobility (a) and diffusion coefficient (b)
versus the tilting force $F$ for $a_0/L = 1.5$ and different
values of damping constant $\gamma$; $T=0.1$ and $K=0.3$.
Two-peaked diffusion curves are clearly distinguishable for
$\gamma \leq 1$ only. } \label{dep-FKg}
\end{center}
\end{figure}

The problem of a dimer diffusing in a washboard potential has been
studied in fact in many papers, but due to the large parameter
space, important effects went unnoticed. In
Ref.~\cite{braun2003nb}, it was found that for a commensurate
dimer, $D(F,T)$ had two maxima as a function of the tilting force
$F$, whereas an incommensurate dimer behaved more like a monomer,
with $D$ showing only one peak. However, as shown in
Fig.~\ref{dep-FK}(b), one can observe two $F$ maxima also in the
diffusion coefficient of a noncommensurate dimer;
correspondingly, the mobility curve $\mu$ versus $F$ develops the
nonmonotonic behavior displayed in Fig.~\ref{dep-FK}(a). More
remarkably, for the same temperature and damping constant of
Fig.~\ref{dep-FK}, commensurate dimers presented a single peaked
diffusion coefficient and monotonic mobility as functions of the
tilt (see Figs.~\ref{dep-FK-a1} and \ref{dep-Fa}). However, for
different simulation parameters (like those in
Ref.~\cite{braun2003nb}) two-peaked $D$ curves were detected for
commensurate dimers, as well. Thus, a doubly peaked diffusion
coefficient is no signature of dimer-substrate commensuration:
the coupling constant (Fig.~\ref{dep-FK}), damping constant
(Fig.~\ref{dep-FKg}), and temperature also play a significant
role [see Eqs.~\ref{F1} and \ref{F2}].

To investigate the origin of the two competing diffusion mechanisms
shown in Fig.~\ref{dep-FK}, we address in detail the case of a dimer
with length $a_0$ equal to a half-integer multiple of the substrate
constant $L$. For a finite coupling strength $K$, on setting
$Y(t)=a_0+\psi(t)$, the coupled LE's (\ref{CMR2a}) and (\ref{CMR2b})
read
\begin{eqnarray}
\ddot{X} &=& - \gamma \dot{X} - \sin{(\psi/2)}\sin{X} + F + Q (t)/\sqrt{2} \, , \label{CMR3a} \\
\ddot{\psi} &=& - \gamma \dot{\psi} + 2\cos{(\psi/2)} \cos{X} - 2K\psi +
\sqrt{2} q (t).~~ \label{CMR3b}
\end{eqnarray}
If the dimer is sufficiently rigid and the tilting force $F$ weak,
then $\psi (t)$ is small and mostly controlled by thermal noise.
From Eq.~(\ref{CMR3b}), on neglecting the substrate force with
respect to the dimer coupling, energy equipartition yields
$\langle \psi^2(t) \rangle =T/K$. Moreover, the force term
$\sin{(\psi/2)}\sin{X}$ in Eq.~(\ref{CMR3a}) can be treated as
resulting from a randomly flashing cosine potential with amplitude
$2\langle |\sin{[\psi(t)/2]}| \rangle \thickapprox |\psi|$. 
This can be regarded as an instance of the ``parametric resonance'' approach
pursued by the authors of Ref.~\cite{fusco2003TSF} in the limit $T=0$.
On assuming a Gaussian distribution for $\psi$, a corresponding
$\gamma$-independent effective critical tilt can thus be
estimated, namely,
\begin{equation}
F_{1} \approx [(2/\pi) \langle \psi ^2(t) \rangle] ^{1/2} =
\sqrt{2T/\pi K}. \label{F1}
\end{equation}
As pointed out in Sec. \ref{tilting}, for large to intermediate values of
damping, the critical tilt coincides with the effective dimer
depinning threshold $F_d$. For $K\geq 0.2$, Eq.~(\ref{F1}) locates
rather accurately the first $F$ peak of the simulated diffusion
coefficient reported in Fig.~\ref{dep-FK}(b).

For $F>F_1$ both the dimer mobility and the diffusion coefficient
tend towards their free particle values, unless an internal
resonance sets in. Indeed, driven by a strong force $F$, the dimer
center of mass acquires an almost constant speed $F/\gamma$. On
inserting $X(t)\simeq Ft/\gamma$ into its right hand side,
Eq.~(\ref{CMR3b}) becomes the LE of a Brownian oscillator
subjected to a harmonic force with angular frequency
$\Omega=F/\gamma$. Accordingly,
the internal degree of freedom of the dimer, represented by the
coordinate $Y$, resonates for $F/\gamma$ approaching
$\sqrt{2K-\gamma^2/2}$ 
(parametric resonance~\cite{braun2003nb,fusco2003TSF,strunz1998,cattuto}), 
thus leading to a threshold-like
enhancement of the dimer diffusion \cite{costantini}. Our argument
can be refined further by noticing that at resonance the processes
$X(t)$ and $\psi(t)$ synchronize their phases, so that the
substrate force in Eq.~(\ref{CMR3a}) does not average out any
more. In the presence of synchronization, $\langle
\sin{(\psi/2)}\sin{X}\rangle \simeq 1/2$, which amounts to
replacing $F$ with $F-1/2$. In conclusion, for relatively large
damping constants, namely $1 \lesssim \gamma <2\sqrt{K}$, a
resonance diffusion $F$ peak is expected for
\begin{equation}
F_{2} \approx \frac{1}{2}+\gamma\sqrt{2K-\frac{\gamma^2}{2}}.
\label{F2}
\end{equation}
in reasonable agreement with the simulation results of
Fig.~\ref{dep-FK}(b) for $\gamma =1 $. Correspondingly, the
mobility curves describe a two step transition from the locked to
the running state.

For weak dimers, $K < (\gamma/2)^2$, the two peaks of the
diffusion coefficient tend to merge, as shown in
Fig.~\ref{dep-FK}(b), and in the limit $K \to 0$ a monomer
dynamics is recovered (see Sec. \ref{tilting}). Equivalently,
incommensurate dimers with $\gamma>2\sqrt{K}$ must be regarded as
overdamped as far as their internal coordinate $Y$ is concerned;
therefore, their diffusion coefficients are characterized by one
maximum located around the $\gamma$-independent depinning
threshold $F_d$ in Eq.~(\ref{F1}); see Fig.~\ref{dep-FKg}(b). When
$\gamma$ decreases, both diffusion peaks shift towards smaller
values of $F$. The explanation is very simple: The resonance
threshold $F_2$ tends almost linearly to $1/2$; in the underdamped
regime, the depinning threshold $F_d$ is proportional to $\gamma$
as it obeys law (\ref{lowg}) with $F_{\mathrm{cr}}$ given by the
effective critical tilt $F_1$ of Eq.~(\ref{F1}). This estimate for
$F_d$ in the underdamped limit is consistent with the anticipated
locked-to-running transition thresholds exhibited by the mobility
curves of Fig.~\ref{dep-FKg}(a) with $\gamma \lesssim 0.3$.

Going back to the dynamics of the damped incommensurate dimer of
Fig.~\ref{dep-FK}, we remark that on increasing $K$ the resonance
diffusion peaks, in addition to shifting to higher $F$ (directly
proportional to $\sqrt{K}$), flatten out on top of the plateau
$D=D_0(T/2)$; as the depinning peaks move to lower $F$ (inversely
proportional to $\sqrt{K}$), for $K\to \infty $ the diffusion
coefficient eventually tends to $D_0(T/2)$, as anticipated in the
previous Sections.

The argument presented here can be easily generalized to the case
of commensurate dimers, or to any equilibrium length; the ensuing
properties of commensurate versus noncommensurate dimers and the
different monomer limits of the dimer dynamics have been
anticipated, respectively, in Secs. \ref{length} and
\ref{tilting}.

\section{Conclusion}
\label{conclusion}

In this paper we have studied a system consisting of two
harmonically interacting Brownian particles diffusing in a 1D
washboard potential. We found that the average current and
the diffusion coefficient of such a dimer exhibit a complicated
non-monotonic behavior as a function of the driving force and the
ratio of the dimer length to substrate constant. In the limits of
the weak ($K \to 0$) and strong ($K \to \infty $) coupling
constant the expected monomer dynamics was recovered. Moreover, we
studied in detail the dimer transport for different coupling
strengths and damping constants. We concluded that the
appearance of the second resonant peak of the diffusion
coefficient versus the driving force is not related to the dimer
length-to-substrate constant ratio, but rather to the
damping-to-coupling constant ratio; the diffusion coefficient
$D(F)$ possesses two peaks only for relatively low damping values.

Finally, we recall that a simple 1D model is not always a viable
tool to analyze transport in two or higher dimensions: such a
modeling makes sense for highly symmetric substrates, only. There
exist irreducible 2D and 3D devices where particles are driven on an
{\it asymmetric} potential landscape by an ac or dc driving force
perpendicularly to the symmetry axis of the potential. Such a
geometry has recently attracted broad interest \cite{2Daltro} in the
context of separation of macromolecules, DNA, or even cells, because
it is capable of inducing a transverse drift as a function of the
drive and of the particle geometry: as a consequence different
objects can be separated depending on their center of mass diffusion
coefficient \cite{2D}. While the motivations of the present study
apply to this class of devices, too, it is clear that their
characterization must take into account the dimensionality of the system
at hand. Dimensional reduction is limited by the spatial symmetry of
the substrate and the particles. This is the subject of ongoing
investigation.

\begin{acknowledgments}
This work has been supported by the Estonian Science Foundation
via Grant No. 6789 and by the Archimedes Foundation (EH).
\end{acknowledgments}


\end{document}